SpaceOps-2025, ID # 428

# Evolution of JWST Contingency Payload Operations for Mitigating NIRSpec Micro-shutter Array Electrical Shorts

Katie Bechtold[a]*, Torsten Böker[b], David E. Franz[c], Dennis Garland[d], Maurice te Plate[e], Timothy D. Rawle[f], Christopher Q. Trinh[g], Rai Wu[h], Peter Zeidler[i]

[a] *Space Telescope Science Institute, Baltimore, USA, kbechtold@stsci.edu*
[b] *European Space Agency (ESA), ESA office at STScI, Baltimore, USA, boeker@stsci.edu*
[c] *NASA Goddard Space Flight Center, Greenbelt, USA, david.e.franz@nasa.gov*
[d] *Space Telescope Science Institute, Baltimore, USA, dgarland@stsci.edu*
[e] *European Space Agency (ESA), ESTEC, Noordwijk, Netherlands, maurice.te.plate@esa.int*
[f] *European Space Agency (ESA), ESAC, Madrid, Spain, tim.rawle@esa.int*
[g] *Space Telescope Science Institute, Baltimore, USA, ctrinh@stsci.edu*
[h] *Space Telescope Science Institute, Baltimore, USA, rwu@stsci.edu*
[i] *AURA for the European Space Agency, ESA Office, STScI, Baltimore, USA, zeidler@stsci.edu*
* Corresponding Author

## Abstract

The Near-Infrared Spectrograph (NIRSpec) is one of four science instruments on board the James Webb Space Telescope (JWST), which began routine operations in July 2022. As JWST's primary spectroscopic instrument for faint, distant targets, NIRSpec plays a central role in several of the mission's core science goals. Its signature multi-object spectroscopy (MOS) mode enables the simultaneous acquisition of spectra for up to a hundred targets across the field of view, using the first space-based micro-shutter array (MSA), comprising nearly 250,000 individually addressable micro-electromechanical shutters. This capability makes it feasible to survey the sparsely distributed population of faint galaxies and study galaxy formation in the early universe using statistically significant samples [1, 2]. The number of usable micro-shutters directly impacts the number of simultaneously observable targets, making the preservation of shutter functionality a priority for the NIRSpec Mission Operations Team.

The MSA is susceptible to occasional electrical shorts, likely due to particulate contamination. These shorts produce unwanted infrared glow in MOS exposures, rendering them unusable and wasting valuable observatory time. Mitigation requires promptly identifying the affected shutter(s) and masking the corresponding row(s) or column(s) to prevent future activation. However, masking shutters unnecessarily reduces NIRSpec's multiplexing capacity, so it is important to minimize the extent of masking while reliably suppressing the short's effects.

Procedures for remotely locating and masking shorts were established in the course of integration and testing during the years leading up to the JWST launch[3]. More than two years of operational experience[4] have informed refinements to these procedures and to the operational decision-making that guides when and how they are applied. This experience also supports new efforts to identify and unmask previously affected rows or columns where shorts have since disappeared, enabling the recovery of multiplexing capacity that would otherwise remain lost.

As micro-shutter array technology matures toward use in future missions such as the Habitable Worlds Observatory, the operational strategies developed for NIRSpec remain relevant beyond JWST. This paper provides an overview of micro-shutter shorts mitigation operations, summarizes shorts observed since the completion of commissioning, and describes the evolving contingency procedures that have enabled continued science productivity from this unique payload.

**Nomenclature**

$V_{PP}$    Positive peak voltage; the high-voltage supply for the HV584 driver chip that sets the output level used to actuate micro-shutter rows or columns

**Acronyms/Abbreviations**

Association of Universities for Research in Astronomy (AURA), Astronomer's Proposal Tools (APT), data number per second (DN/s), Deep Space Network (DSN), electrical short detection (ESD), flight systems engineer (FSE), integral field unit (IFU), integrated science instrument module (ISIM), James Webb Space Telescope (JWST), micro-shutter array (MSA), micro-shutter control electronics (MCE), Near-Infrared Spectrograph (NIRSpec), Observation

Plan (OP), optical short detection (ESD), project reference database (PRD), Space Telescope Science Institute (STScI), tri-state mask (TSM), zero-potential mask (ZPM)

## 1. Introduction

*1.1 NIRSpec and the micro-shutter array*

The Near-Infrared Spectrograph (NIRSpec) is one of JWST's four science instruments and supports multiple observing modes, including single-object, integral field, and multi-object spectroscopy (MOS). MOS is achieved using a programmable slit mask known as the Micro-Shutter Array (MSA), which enables the simultaneous acquisition of spectra for tens to hundreds of astronomical targets within a single exposure. This approach is not feasible with traditional single-slit spectrographs, which are limited to one target at a time and require mechanical reconfiguration between pointings.

On ground-based telescopes, MOS is typically implemented using custom-machined slit masks or robotically positioned optical fibers. However, the mass, volume, and complexity of such systems make them impractical for spaceflight. To provide aperture flexibility within a compact, lightweight footprint, NIRSpec uses a micro-electro-mechanical system (MEMS) known as the Micro-Shutter Array (MSA). The MSA consists of four quadrants, each mounted on a cruciform support structure and comprising a rectangular grid of microscopic shutters: 171 in one direction and 365 in the other. Each shutter, measuring approximately 78×178μm, rotates 90° on a flexure hinge to open or close the light path to the spectrograph.

Programming these shutters is accomplished by a combination of electrostatic addressing and magnetic actuation. The shutter doors open as their magnetic coating reacts to a transient field produced by a permanent magnet mounted on a "magnet arm" that sweeps laterally over the shutter array. To hold certain shutters open, an electrostatic field is created by applying a voltage differential between electrodes on the shutter doors and electrodes on the support grid walls. When the voltage differential is removed, torsion in the flexure hinge causes the shutter to spring closed.

NIRSpec's micro-shutter control electronics (MCE) applies these electrostatic fields via a set of cryogenic, radiation-hardened HV584 high voltage drivers mounted on the substrate surrounding each quadrant. One subset of each quadrant's drivers controls the negative voltage rails connecting the electrodes on the shutter doors, while another subset controls the positive voltage rails connecting the electrodes on the grid walls. Electrically, each shutter functions as a potential intersection between one positive and one negative line.

Shutter pattern configuration is accomplished in three steps. First, all shutters are latched open as the magnet arm traverses the array from its nominal rest position (called Primary Park) to the other side of the array. Next, the desired aperture pattern is established by synchronizing the shutter voltages with the magnet's return traversal so that as the magnet passes over the array again to return to Primary Park, all shutters except those included in the desired pattern are unlatched, causing them to close. Finally, once the array has been configured and the magnet has returned to Primary Park, the remaining latched shutters are held in place at a quiescent voltage. The procedure is repeated for each shutter pattern desired in an observation, and when the observation is complete, the procedure is repeated a last time with all shutters released to close the entire MSA aperture when not in use.

*1.2 JWST observation structure and onboard execution*

The James Webb Space Telescope is designed for autonomous operation. Observations are planned on the ground and defined in terms of visits, which specify sequences of spacecraft and instrument activities. These visits are assembled into an Observation Plan (OP), which is uploaded to the observatory during scheduled communications via the Deep Space Network (DSN). Onboard, the Observation Plan Executive (OPE) manages the execution of the OP by dispatching visit-level command sequences to the spacecraft and instruments. Within each visit, the Operations Scripts Subsystem (OSS) interprets and executes lower-level command scripts that carry out the activities defined in the visit structure[5,6]. Real-time commanding beyond routine activities is typically limited to specific, carefully coordinated DSN contacts.

Observatory time is a highly valuable and limited resource, and JWST's observation planning is designed to maximize science return. Observations are selected during annual proposal cycles and executed as part of approved science programs. Instruments may operate in parallel—such as NIRCam imaging while NIRSpec performs spectroscopy—allowing more efficient use of telescope time.

Observation configuration, including aperture selection, is defined in advance. Relevant resource status information such as shutter masks are stored on the spacecraft in version-controlled files drawn from JWST's official Project Reference Database (PRD). Updating these files onboard follows a tightly controlled process: the change must be certified via ground testing with a flight software simulator, then reviewed and approved by a change control board before uplink. Certified changes are incorporated into the OP package for future execution.

Some operational activities are executed by dedicated onboard scripts, while others—typically time-critical or diagnostic tasks—may require real-time commanding during a scheduled DSN contact. Exposure data are stored onboard and later downlinked during scheduled DSN contacts. Once received on the ground, the data are processed into calibrated science products, distributed to the original observers, and after a proprietary period, released to the broader scientific community. If an observer identifies a problem that compromised the execution of a visit—such as instrument configuration issues or anomalies not caused by the target or proposal—they may request that the observation be repeated, subject to review and approval by the mission.

These constraints and tools define the operational landscape in which NIRSpec short detection and mitigation must be planned and executed.

**2. MSA electrical shorts**

Like any MEMS device, the NIRSpec MSA is susceptible to intermittent short circuits. In general, a short on a MEMS device can result from contamination, mechanical deformation, oxidation, or line deformation, but particulate contamination is believed to be the only plausible source of new NIRSpec MSA shorts in space flight. Under cryogenic conditions, even small current elevations can cause a temperature increase of the shutter grid, and the associated infrared emission can contaminate astronomical data sets. In severe cases, this can even cause physical damage to the quadrant circuits or MCE.

*2.1 Impact of shorts*

While none of the MSA shorts detected in the first three years of science operations have produced current levels that pose a danger to the hardware, most of them caused significant contamination of science data. Some shorts produce only a faint glow on a small portion of the field of view. Other shorts saturate significant portions of the detectors with glow not only from the point of contact on the array but also from reflections of that glow on the shutter grid, array frame, integral field unit (IFU) aperture, and other structural elements. In such cases the exposure is rendered useless for science. Fig. 1 illustrates the effect of a new short, in contrast to an equivalent version of that exposure without the short, which is shown in Fig. 2. In addition, under certain conditions glow from a short on the MSA can radiate and reflect back through the NIRSpec fore optics and pick-off mirrors, reflect off the telescope's secondary mirror, and enter the light path of another instrument exposing in parallel with NIRSpec. This phenomenon is believed to be the cause of anomalous effects in exposures like the NIRCam one shown in Fig. 3, taken when NIRCam was exposing in parallel with NIRSpec.

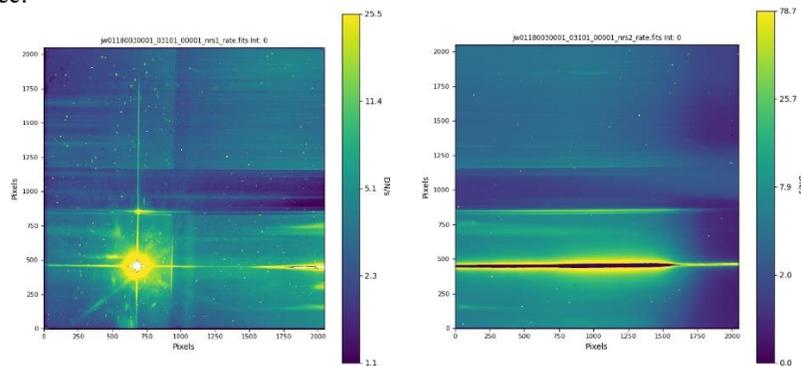

Fig. 1. A pair of 2-D count rate images from a science exposure exhibit glow from an October 2022 short dispersed on the two NIRSpec detectors. (Note that saturated pixels are displayed with a count rate of 0 DN/s.) This exposure was part of an 11.46-hour observation for JWST Proposal 1180[7] which was granted a repeat due to unrecoverable data loss from the contaminating glow. The short was masked 17 days after this observation.

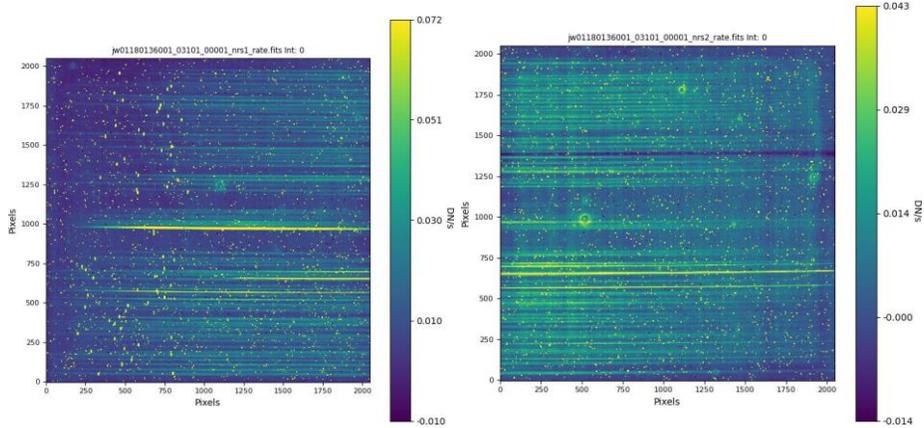

Fig. 2. A successful October 2023 repeat of the exposure shown in Fig. 1

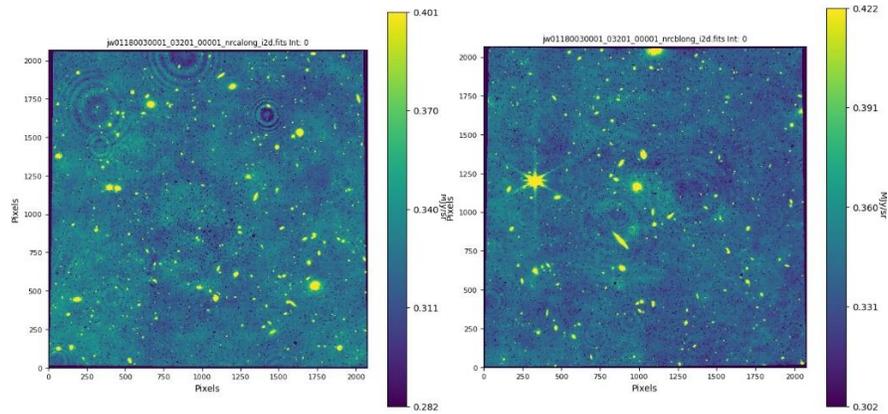

Fig. 3. Radiation from the short illustrated in Fig. 1 also affected NIRCam exposures that were taken in parallel with the NIRSpec visit. Shown here are NIRCam's two long-wavelength detectors, illuminated by an unusual ripple pattern that is likely caused by glow from the MSA short being reflected into the NIRCam optics.

Observations affected by spectral contamination from MSA shorts typically require repetition in order to meet their science goals, which results in a direct cost to observatory efficiency. As of February 2025, a total of 169.6 hours of JWST observing time have been spent repeating science observations due to short-induced contamination (see middle subplot of Fig. 4). In addition, 15.1 hours have been devoted to engineering visits for short detection (see bottom subplot of Fig. 4). While this investment has ensured the scientific usability of affected programs, it has also displaced other potential observations—highlighting the importance of proactive mitigation.

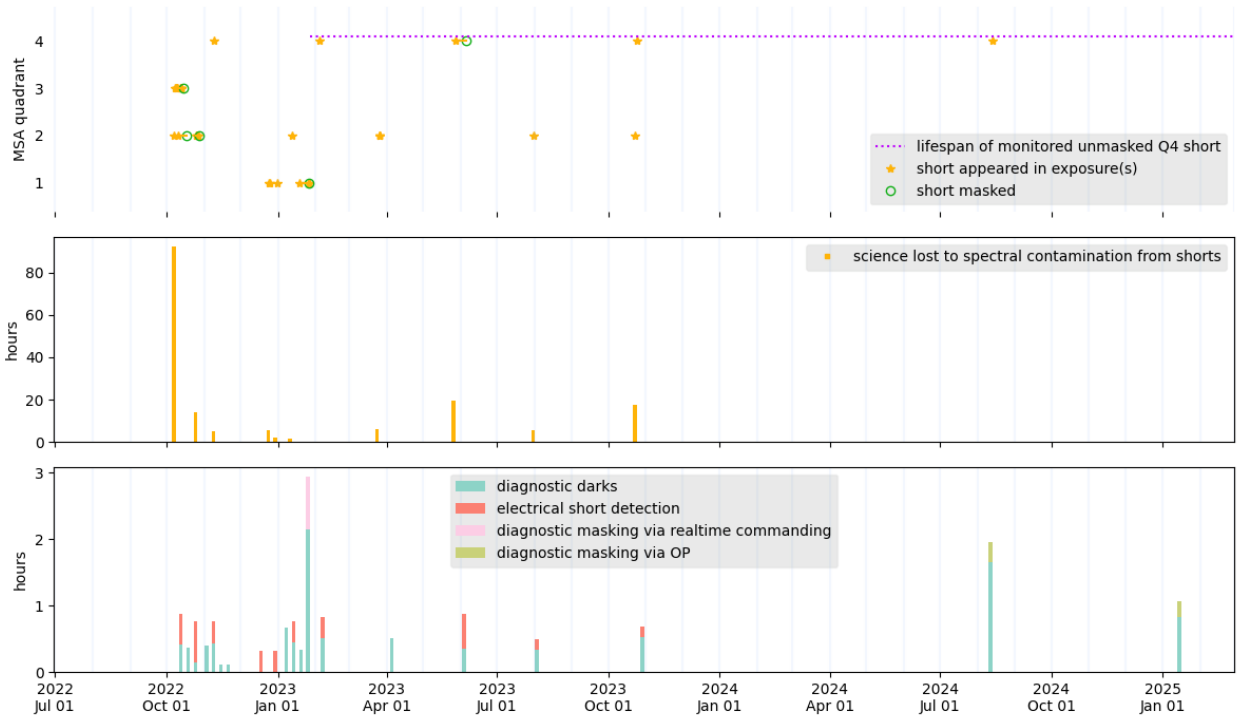

Fig. 4. Timeline of post-commissioning MSA shorts. Top: appearances of shorts presented chronologically by MSA quadrant, calling out occasions when shorts were masked along with the lifespan to date of a chronic low-level short on Quadrant 4. Middle: bar chart depicting the timeline of 'hours lost' per four-day interval, as measured by observation time approved to be repeated based on contamination from an MSA short. Bottom: bar chart showing hours of observatory time expended in various activities relating to short detection. Note that the January 2025 activities reflect a re-checking of previously masked shorts as described in Section 6.

*2.2 Types of shorts*

Electrical shorts on the micro-shutter array arise from unintended current paths between components such as voltage rails, shutter electrodes, and the supporting substrate (see Fig. 4). Shorts involving current flow between adjacent voltage rail lines on the same side of the array are referred to as "nearest-neighbor" shorts; see Section 4.3 for discussion of recent improvements to onboard scripts that enhance detection of this short type. In contrast, "front-to-back" shorts occur when the current path bridges the front-side circuit connected to shutter-door electrodes and the back-side circuit connected to the grid-wall electrodes. Other unwanted current paths may involve the array or quadrant substrates themselves.

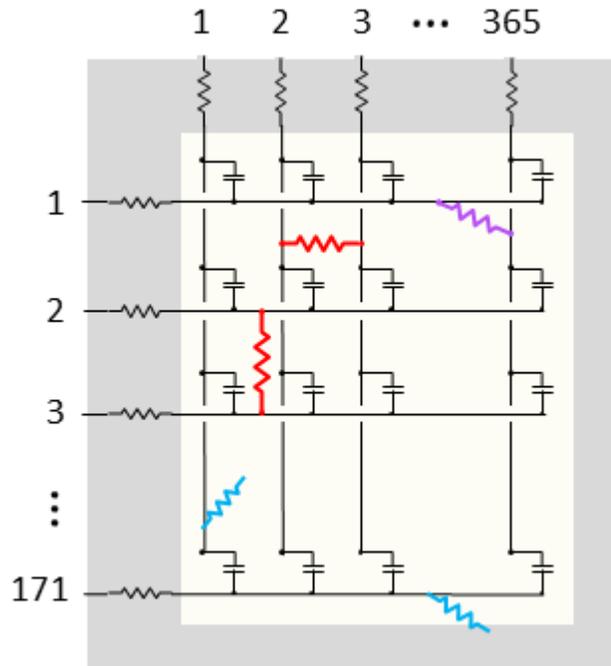

Fig. 4. Array Short Scenarios. A simplified circuit schematic depicts a micro-shutter array as a multiplexed 365×171 grid of parallel plate capacitors. Three different types of potential short circuit paths are shown: the red paths conduct current between adjacent rails, termed a "nearest-neighbor" short; the purple path conducts current between a 171-side rail and a 365-side rail, described as a "front-to-back" short; and the blue paths conduct current between a driver output rail and the array or quadrant substrate.

*2.3 Short masks*

Short circuits on the MSA can be mitigated by masking the 365- or 171-side line(s) aligned with the point of unwanted electrical contact using two slightly different masking techniques.

One type, a zero-potential mask (ZPM), overrides any shutter pattern data that addresses the masked line(s) during the magnet arm's programming sweep, inhibiting the application of a voltage differential and so preventing the flow of current via the unintended path of the short circuit. To apply ZPMs, before loading the shutter pattern data to the HV584 driver chips, the flight software removes any ZPM-masked lines from the set of open shutters in the pattern data.

The other type of short mask, referred to as a tri-state mask (TSM), removes masked lines from the circuit. In contrast to the ZPM approach, the HV584 driver chips themselves use the TSMs stored in their mask registers to control the high voltage output. When they disable output to a line, the corresponding shutter elements are disconnected from the circuit and can eventually charge up to the point of becoming stuck in the open position. Because unintentionally open shutters produce spectra that contaminate overlapping target spectra, it is important to minimize the potential for failed open shutters, and so shutter lines containing shorts are masked with ZPMs when possible and only masked with TSMs as a last resort.

The JWST integrated science instrument module (ISIM) flight software maintains the ZPMs and TSMs in text files within an instrument-module-wide RAM filestore, based on version-controlled mask files in JWST's official project reference database (PRD) for data loaded to the observatory. In addition, the operative ZPMs reside in MCE

RAM, and the operative TSMs reside in mask registers in the HV584 driver chips; these are the versions that are actually applied to the MSA, and they are only updated when commanding is needed to load them to MCE RAM from the ISIM-wide RAM. The full set of 16 masks includes masks for every combination of quadrant, side (171 or 365), and mask type. Updating the onboard masks follows a tightly controlled process requiring ground certification testing with a flight software simulator followed by approval of the activity and test results by a change control board prior to uplink. The updated masks are then included as entries in the PRD.

The updated short masks are then propagated to the MSA operability map within the JWST Astronomer's Proposal Tools (APT) observation planning software so that science users can plan their observations with up-to-date maps of usable shutters. This process needs to happen in a timely manner in order to avoid an unexpected loss of science targets when the mask update has added to the population of inoperable shutters.

### 3. Operational short mitigation

*3.1 Electrical short detection*

The NIRSpec team's standard first response to a newly identified short is to attempt to locate and mask it. This is achieved via an activity called electrical short detection (ESD) and is performed by a dedicated on-board script. The script can be invoked either via real-time commanding or as part of an Observation Plan (OP) package[6], i.e. the sequence of autonomously executed JWST activities.

The ESD script detects and isolates electrical shorts within an MSA quadrant by progressively testing shutter line subsets under different masking configurations.

Following an initial diagnostic phase, the script begins the narrowing process using a zero-potential mask (ZPM). At the start of this phase, all lines are masked by default, and the script temporarily unmasks only those not already masked due to previous detections, and the resulting quadrant current is measured. If the current exceeds a user-defined threshold, the script recursively narrows the search by testing progressively smaller subsets to isolate specific shorted lines. In some detection phases, however, the script defers narrowing until broader diagnostic checks are complete.

If the quadrant current does not exceed the threshold with the initial ZPM, the script proceeds to evaluate an alternate masking strategy using a tri-state mask (TSM). If none of these configurations are effective in isolating the short, the algorithm terminates.

When one or more shorted lines are identified by the narrowing process, they are flagged for inclusion in the on-board short mask corresponding to the mask type used during detection. The script continues evaluating the remaining lines under the same masking approach to ensure that any additional shorts are also identified and flagged. The script concludes by updating the ZPMs to include any lines flagged as shorted. TSMs are not modified by the script.

After this on-board script has completed, if it has updated any on-board ZPMs, NIRSpec FSEs coordinate the configuration control processes to incorporate the update into the PRD.

This systematic process supports the reliable detection and masking of shorts while minimizing disruption to science operations. Section 4.3 describes an in-flight update to improve the detection of nearest-neighbor shorts, a category that was often missed by earlier implementations.

When a new short appears, electrical stimulation and discharge (ESD) is scheduled as soon as practical. During commissioning, ESD was often performed via real-time commanding. In routine operations, however, autonomous execution as part of the onboard Observation Plan (OP) is strongly preferred due to the logistical constraints of coordinating real-time commanding. OP-based ESD execution involves creating an ESD ``visit'' in the Astronomer's Proposal Tool (APT), specifying the affected quadrant(s) and threshold parameters. This visit is typically paired with an internal diagnostic exposure designed to determine whether the short has persisted following the ESD execution.

*3.2 Optical short detection*

In some cases, a short that continues to produce glow cannot be isolated by ESD, such as when the short does not raise the current above the detection threshold, or prior to deployment of the nearest-neighbor short location functionality described in Section 4.3 when a nearest-neighbor short appeared. In these cases the team can attempt optical short detection (OSD). This process involves testing candidate masks on-board to determine which effectively, and most efficiently, masks the short.

The first step in this process is identification of several likely lines as the short's location based on the diagnostic exposures that will already have been taken in conjunction with the unsuccessful ESD. For this step, the NIRSpec instrument model can generally be used to infer the short's position to within ±1 line in each direction (side).

Once candidate lines are identified, a set of corresponding candidate masks is generated. These candidate masks are for one-time use in the detection activity and are not incorporated into the PRD. Nonetheless, since they will be used with flight hardware they are tested with the same process used for official, certified PRD items.

Once the candidate masks have been tested, the NIRSpec FSE team uses APT to create a sequence of visits structured to alternate between candidate mask application and corresponding diagnostic exposure, with the pre-existing mask from the ISIM-wide filestore restored at the end of the activity. Because exposures can only be taken by autonomously executed visits in the OP, and because candidate masks could only be applied via ground commanding during real-time visits in early science operations, the activity could not be performed entirely autonomously. As a result, OSD at that stage of operations required a complex series of hand-offs of instrument commanding control—between ground-based controllers for mask application and on-board scripts invoked in the OP for exposures. This constraint was later resolved with the introduction of OP-based masking, as described in Section 4.4.

Once the exposure data is downlinked and processed, offline analysis determines which, if any, of the candidate masks successfully suppressed the short. At this point the chosen mask is reproduced in multiple formats—text files to be maintained in the ISIM-wide filestore, command sequences used to restore the short masks as part of safing recovery, and a shutter map file to be incorporated into the operability map used in APT[8].

## 4. Evolution of short mitigation

The procedures and capabilities for mitigating MSA shorts have evolved in significant ways since JWST commissioning. This section outlines how the NIRSpec team's approach to short detection, localization, and masking was shaped by in-flight experiences, operational constraints, and lessons learned. It describes the transition from early reliance on real-time commanding and manual diagnostics to more autonomous and efficient on-board methods, developed in response to shifting resources and evolving mission practices. The timeline begins at the conclusion of commissioning in mid-2022 and extends through the deployment of OP-based short detection in late 2023, culminating in the early 2025 initiation of an effort to re-check previously masked shorts—an effort that is expected to enable the recovery of hundreds of shutters to full operability.

*4.1 Commissioning the MSA*

After the launch of JWST on 25 December 2021, numerous NIRSpec commissioning activities were planned, executed, and analyzed by ESA and AURA instrument scientists, supported by subject matter experts along with flight systems engineers (FSEs) at the Space Telescope Science Institute (STScI), which encompasses the center of JWST science and mission operations. The first NIRSpec commissioning activities involving the MSA included quadrant and transport mechanism motor power-on, MCE initialization, release of the magnet arm from the lock that secured it during launch, and shorts detection and masking[9]. Performance and calibration of the elements required for NIRSpec MOS, including the MSA, were evaluated and calibrated, including an assessment of MSA operability[10]. At the completion of the JWST commissioning phase NIRSpec MOS was declared ready for science operations on 1 July 2022.

*4.2 Monitoring for new shorts*

In the JWST commissioning phase and the first year of science operations, the unpredictability and potential impact of in-flight MSA shorts drew particular attention, particularly with the MSA operating for the first time in a microgravity environment. In response to these concerns, a two-pronged approach to monitoring for new MSA shorts was implemented, with the goal of minimizing the delay between a new short degrading or ruining science exposures and it being mitigated.

The earliest indication of a suspected short typically appears in the MSA quadrant current readings that are downlinked in the engineering telemetry, which is available on the ground with relatively little delay compared to the more voluminous and intensively-processed science data. At the beginning of science operations in mid-2022, the NIRSpec FSE team added to their regimen of daily engineering telemetry monitoring a sophisticated display of MSA quadrant currents, so that a new short can be recognized, located, and masked promptly.

Fig. 5 and Fig. 6 illustrate typical quadrant current profiles for an MSA shutter reconfiguration with and without an unmasked short present. The magnet arm begins in its Primary Park position while a low ``hold'' voltage is applied to the rail lines, producing a correspondingly low current. When the arm sweeps across the micro-shutters, a higher "open" voltage is applied to latch all of the shutters, producing a higher current level. On the arm's return traversal, the highest "close" voltage is applied to selectively unlatch the shutters that are to be closed in the desired pattern, further raising the current level. Once the arm returns to its Primary Park position, the "hold" voltage—and therefore low current—is again applied. Under nominal conditions, the current at each phase is proportional with the applied voltage. The signature of an unmasked short is a sustained quadrant current elevation significantly above the normal quiescent level beginning soon after completion of the sweep.

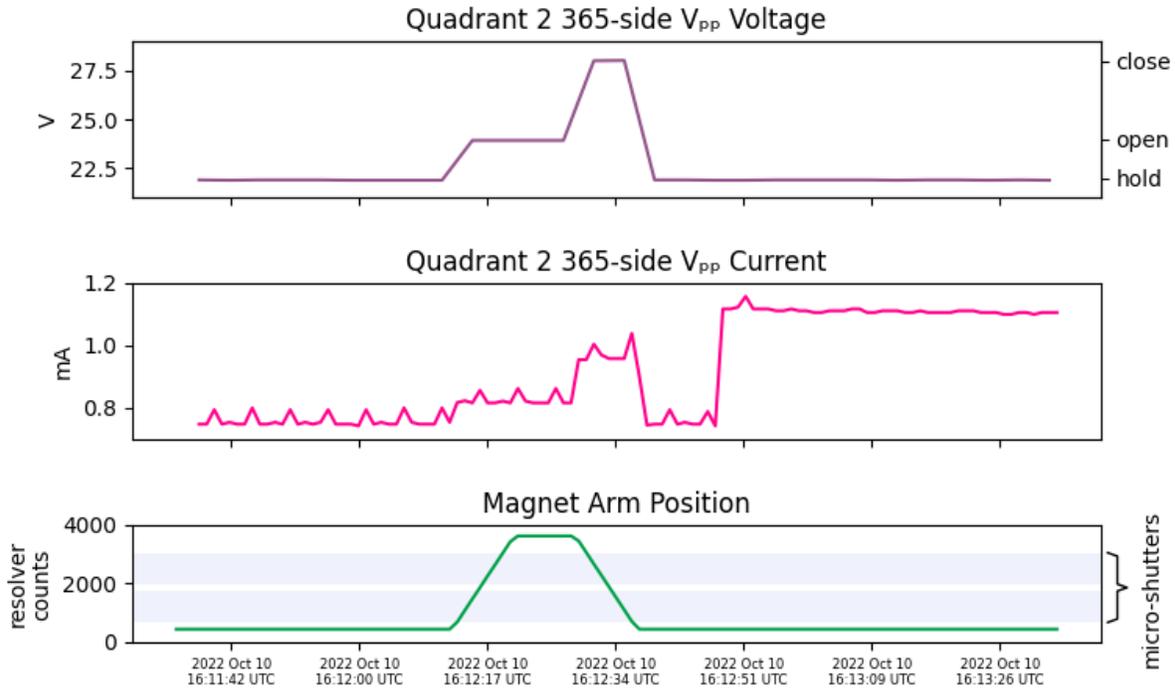

Fig. 5. Quadrant 2 365-side $V_{PP}$ voltage and current, shown with the physical position of the magnet arm on its sweep over the arrays as shutters are synchronously programmed, in this case configuring for the exposure shown in Fig. 1, exhibiting sustained current elevation characteristic of a short circuit

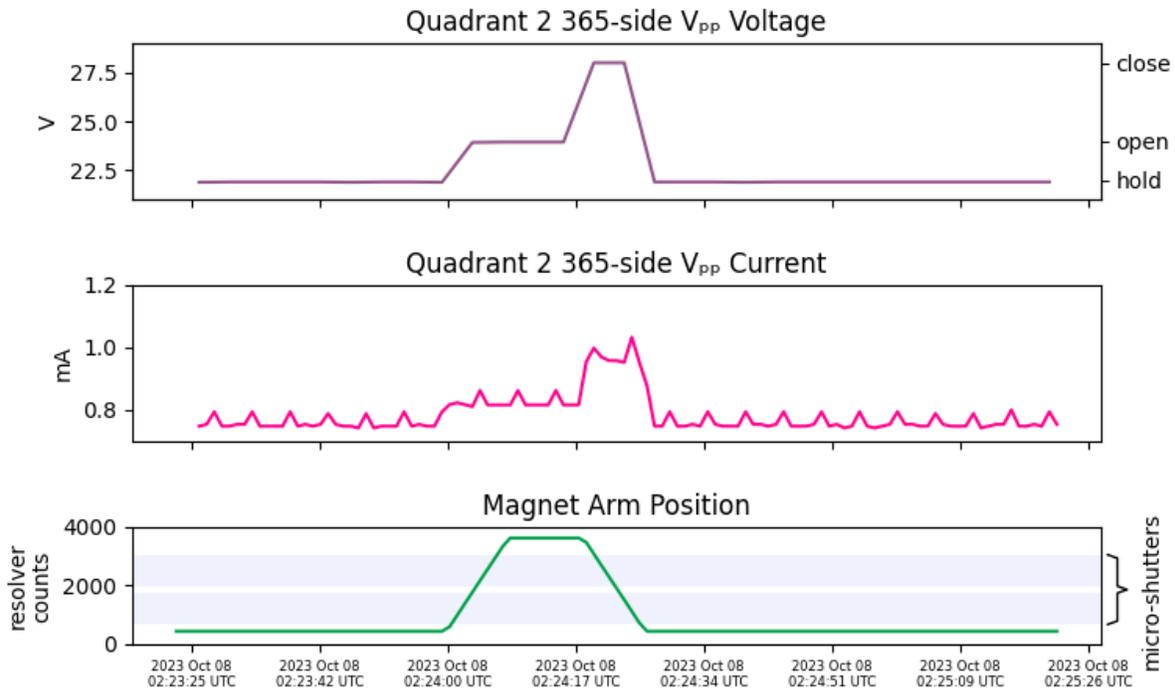

Fig. 6. Quadrant 2 365-side $V_{PP}$ voltage and current, shown with the physical position of the magnet arm on its sweep over the arrays as shutters are synchronously programmed, in this case configuring for the exposure shown in Fig. 2 (a successful repeat of the observation attempted in Fig. 1), exhibiting no sustained current elevation

Although very rare, some shorts may not produce markedly elevated quadrant current readings while still producing unwanted glow in exposures. For this reason, monitoring of the engineering telemetry cannot ensure all shorts are recognized, and so the other element of the short monitoring strategy is review of recent exposures. During instrument commissioning and the first two years of science operations, the NIRSpec instrument science team's rotating role of quicklook analyst entailed responsibility for regular monitoring of exposures' count rate images to flag optical anomalies including ``snowballs"[11], stray light, and glow from new MSA shorts. The goal of this monitoring and flagging was timely recognition of potential issues so they could be characterized, managed when possible in the science data processing pipeline, and communicated to observers. This effort worked in concert with the FSE team's telemetry monitoring until being discontinued in 2024, while the FSE team continues to monitor MSA quadrant current telemetry daily.

*4.3 Nearest-neighbor short location*

The shorts that appeared during commissioning and early science observations often required multiple attempts to locate, with the first and easiest option, ESD, failing to isolate a latent short, and OSD as a more cumbersome fallback option as described in Section 3.2.

One shortcoming of the original on-board ESD script's implementation was its inability to locate nearest-neighbor shorts. Since this type of short involves an inadvertent contact between adjacent lines allowing current on one line to find a path to ground through the neighboring line, activating it requires a voltage differential on one line without a voltage differential on the other side of the point of contact.

However, whenever the original ESD script checked a contiguous set of shutter lines without detecting any currents above the threshold, it did not continue checking successively smaller subsets of shutter lines, on the assumption that the larger set did not contain a short. Soon after this shortcoming was recognized, a straightforward solution was identified to enable the script to detect nearest-neighbor shorts, namely the addition a detecting step in which the set of all even-numbered lines and the set of all odd-numbered lines on each side are checked in the algorithm's initial confirmation that the array as a whole contains one or more locatable unmasked shorts.

Development of this update began six months into normal science operations and was complete and deployed five months later. Coincidentally, a new short appeared three days before that deployment, and execution of the new version of ESD a few days later found it was a nearest-neighbor short that likely would not have been found with the previous version, demonstrating the update's effectiveness in a timely manner.

*4.4 OP-based masking*

Throughout commissioning and the first year of normal operations, the only means of applying candidate short masks on-board for OSD required performing ground-based real-time commanding. However, in the mission's transition from commissioning to science operations, shifts in JWST flight operations resources and practices made this mode of short detection more cumbersome.

Whereas the flight operations team had the benefit of nearly around-the-clock DSN communication with the observatory for continuous real-time commanding during commissioning, JWST's typical DSN coverage currently averages around ten hours per day, limiting real-time commanding opportunities in general. In addition, as a matter of policy since the first year of science operations, non-routine real-time commanding is scheduled to take place only within a few dedicated DSN contacts each month when the observatory is not simultaneously performing science observations. These increased constraints on real-time commanding led the NIRSpec team to prioritize development of a means of performing OSD without the need for real-time commanding.

In February 2023, coordination began to implement the upload and trial application of test masks as a part of the autonomously executed OP that is uplinked on a weekly basis, so that mask trials could be run entirely by on-board scripts, obviating OSD's need for real-time commanding. This effort required extensive coordination among different segments of JWST science operations, including APT, back-end visit scheduling and OP-generating software and databases, and on-board scripts. After the completion of development, testing, reviews, and approvals, the capability was fully deployed in December 2023.

Since the deployment of this OP-based masking capability, no persistent new shorts have appeared, so neither ESD nor OP-based OSD have been executed on a contingency basis since then. However, the NIRSpec team did take advantage of the new capability to locate a persistent low-level Quadrant 4 short that first appeared in January 2023 but has thus far not produced glow bright enough to significantly contaminate science exposures and so has been deliberately left unmasked to avoid sacrificing operability.

Even with the more-convenient OP-based masking technique, OSD still requires the creation and validation of candidate short mask files prior to short detection, and hence OP-based OSD still requires more up-front time and effort than does ESD. Nevertheless, for shorts that can only be detected via OSD, this process has proven to be a significant improvement, saving roughly 25 minutes of execution time compared to the equivalent activity using real-time commanding to apply masks.

The development of OP-based masking also enabled a new operational avenue: revisiting shorts that had been masked earlier in the mission. Although the idea of periodically re-evaluating these shorts had been discussed since before launch, it had remained out of reach due to the complexity of implementing test masks in real time. With autonomous mask trials now integrated into the OP-based visit framework, the team began exploring whether some previously masked shorts might have resolved on their own. This concept—recovering shutter operability through deliberate re-checking—is explored further in Section 6.

### 5. Present operability

Since launch and commissioning, the rate of new short appearances has decreased markedly, as Fig. 4 illustrates, with the most recent new (but transient) short appearing in August 2024. Increases in the number of shutters in short masks, plotted over time in Fig. 7, has likewise plateaued, as the most recent masking of a short took place in June 2023. As a result, degradation of the MSA multiplexing capabilities since that time have been limited to updates in the populations of failed closed and failed open shutters, which have not seen dramatic changes. The operability map at the time of writing is visualized in Fig. 8.

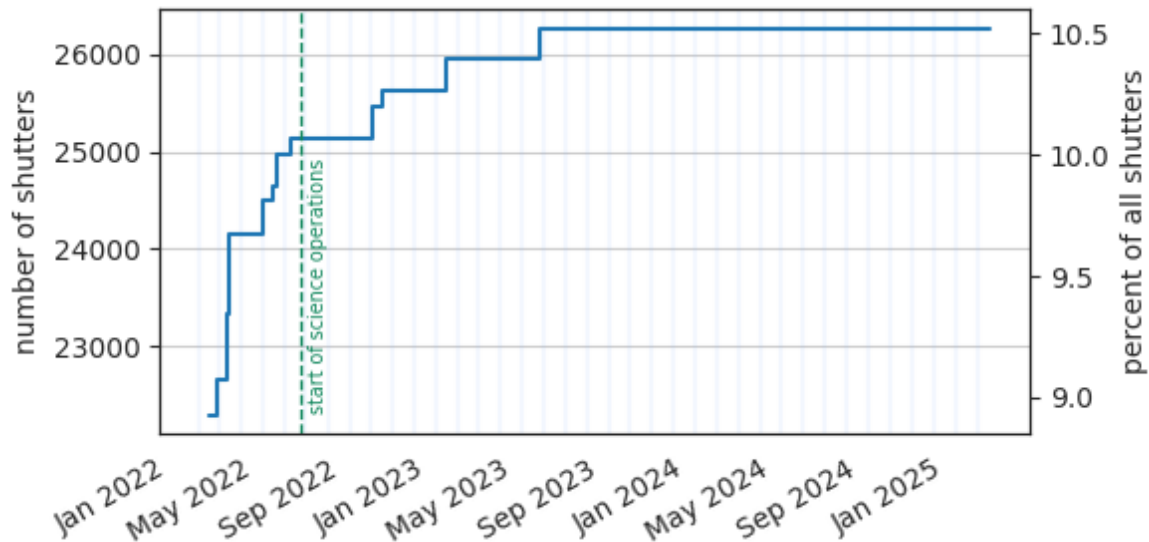

Fig. 7. Total population of shutters in short masks over time, beginning with the first short mask constructed in flight. This population includes shutters that are inoperable for additional reasons beyond their inclusion in a short mask—namely, shutters that are stuck closed or vignetted by a field stop in the NIRSpec fore optics.

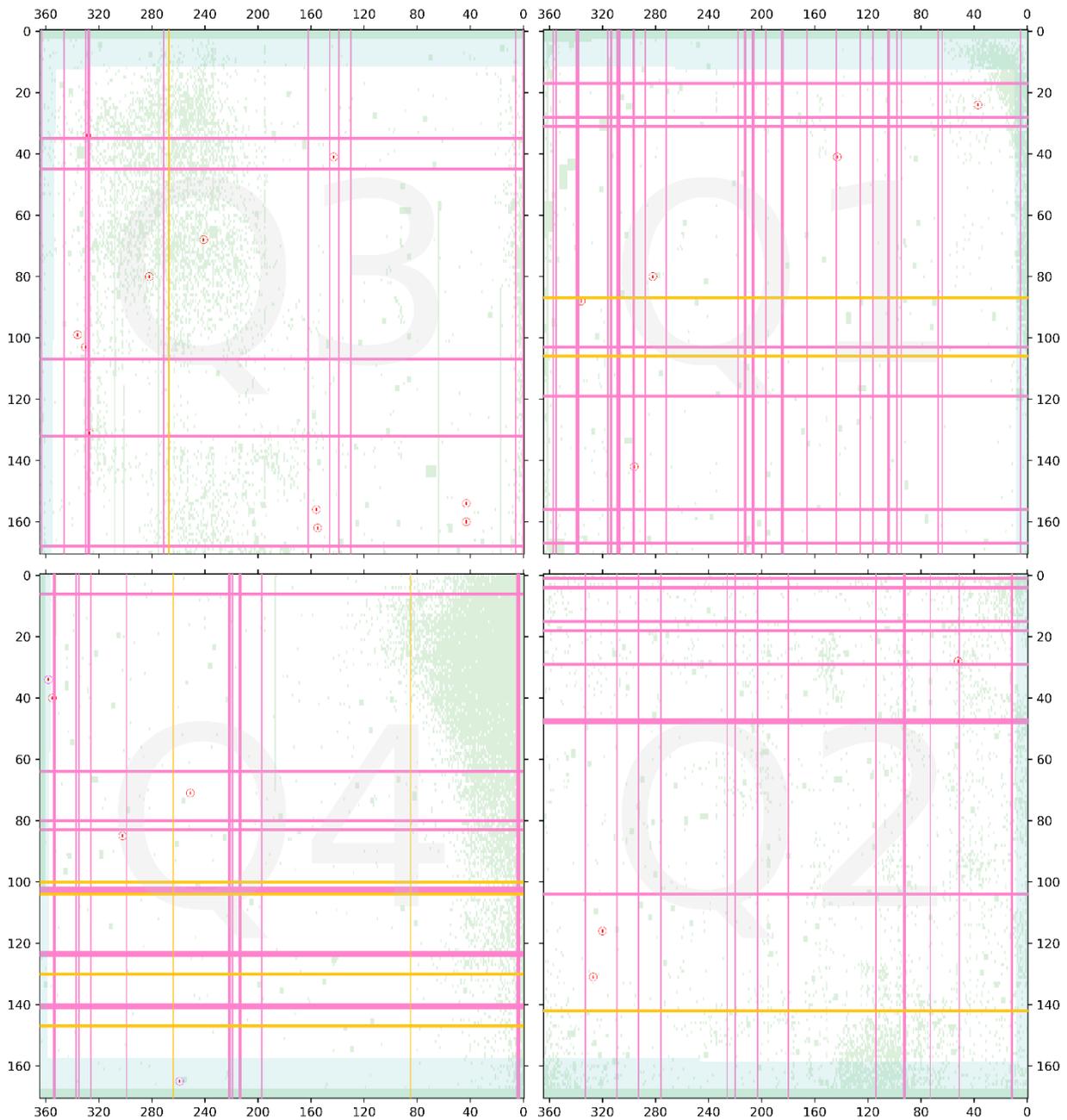

Fig. 8. This visualization of the MSA operability map emphasizes the shutter rows and columns included in the various short masks. Pink lines denote zero potential masks, orange lines represent tri-state masks, light green represents shutters that are stuck closed, light blue represents shutters vignetted by a field stop in the NIRSpec fore optics, and circled red locations represent shutters that are stuck open.

*5.1 A persistent low-level short*

Weak glow from a new Quadrant 4 short was first observed in January 2023 and has since appeared sporadically in the course of numerous multiple observations. Initially believed to be undetectable in quadrant current engineering telemetry, this short was later recognized in February 2025 to correspond with subtle elevations in Q4 171RTN current when certain MSA patterns are applied. These current increases have ranged from 0.5 to 13μA above nominal quiescent levels, a trend that had gone unnoticed until then.

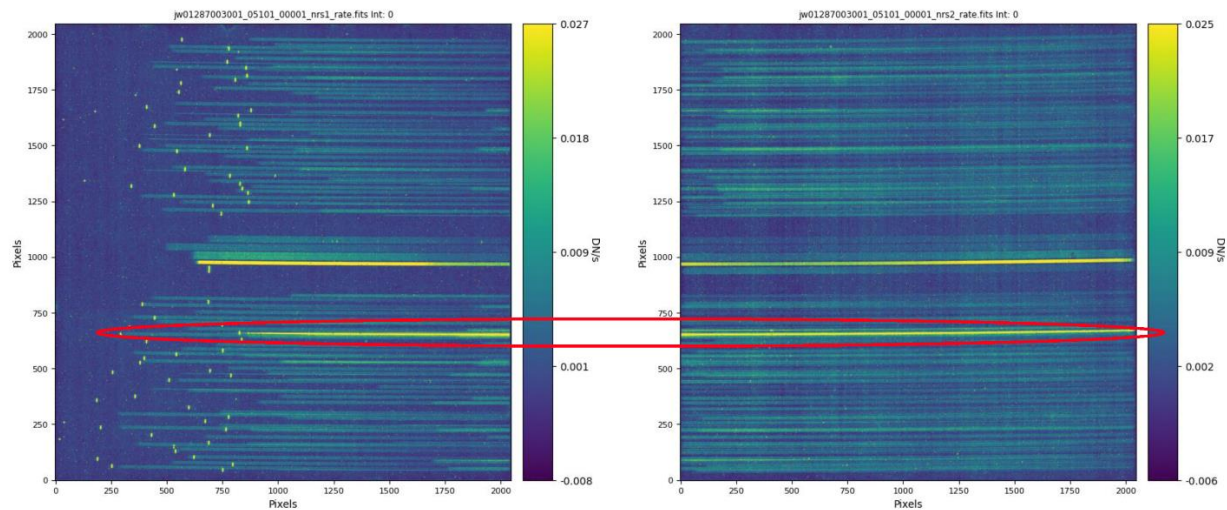

Fig. 9. Glow from a persistent unmasked Quadrant 4 short (highlighted in red) contaminating spectra in JWST Program 1287, Observation 1, executed January 2024

To support future decisions regarding masking and notification, the NIRSpec FSE and instrument science teams conducted a preemptive optical localization of the short in August 2024. A sequence of diagnostic exposures was obtained using internal lamp illumination and custom shutter masks designed to isolate specific row and column intersections on the micro-shutter array. The visit was executed autonomously via the OP and included multiple exposure patterns to confirm the short's position and consistency. The resulting data successfully localized the persistent short to a specific intersection in Quadrant 4 and confirmed that its optical glow matched the feature seen in user-reported observations.

In the course of this activity, a second short with much stronger glow appeared at a different location in exactly one diagnostic exposure. Because this feature did not recur in unrelated calibration exposures using the same shutter pattern the following day, it was concluded to be a transient short. As such, no masking was attempted, and its precise location could not be determined from a single appearance. This incident underscores both the value of timely diagnostics and the challenge of responding to short-lived events whose operational impact is difficult to assess.

6. **Short rechecking**

The prospect of rechecking previously masked shorts with the aim of identifying ones that have since dissipated—which is to say, lines that can be removed from short masks—has attracted interest since years before launch[3]. The convenience of the OP-based masking capability introduced in December 2023 and described in Section 4.4 allowed the NIRSpec team to actively consider performing short re-checking on a discretionary—as opposed to contingent—basis. The inaugural execution of short re-checking was performed in January 2025 using OP-based masking to determine whether two previously-masked shorts were still present. This re-check found that one of the shorts had disappeared at some point since it was originally masked. At the time of writing, the process of updating the Quadrant 4 365-side ZPM to remove the neighboring pair of lines that have been masking it is underway, with the expectation that approximately 288 of those previously-masked shutters can be recovered to full operability (some of the masked shutters are also vignetted and/or failed closed and will not be recoverable).

On the heels of this success, a robust program of re-checking previously masked shorts is being planned for execution over the course of the next one or two years. Although MSA multiplexing capacity is not substantially compromised by the number of inoperable shutters at present, it has been noted that the ability for NIRSpec to perform target acquisition for visits using the MSA is more sensitive to losses of shutter operability[12]. Ultimately, recovery of shutter operability stands to benefit MOS science productivity on both counts.

7. **Conclusions**

Approaching three years into science operations, the NIRSpec team has gained invaluable experience operating the first ever space-based micro-shutter array for multi-object spectroscopy. The unpredictable short circuits that proved a bugbear in early operations have become less formidable not only as new shorts have become less frequent but also as the NIRSpec team has learned and improved its proficiency in responding to them. As development of micro-shutter array technology advances in anticipation of its use on the Habitable Worlds Observatory (HWO), it is hoped that lessons learned from NIRSpec operations can inform future approaches to managing operability. One such effort, the Next-Generation Micro-Shutter Array being developed for HWO, builds on JWST heritage by replacing magnetic actuation with simplified all-electrostatic control and incorporating improved insulation and shutter geometry[13]. While these changes aim to reduce the risk of electrical shorts, the NIRSpec experience underscores the continued importance of diagnostic capabilities and adaptable operations concepts in future architectures.

With fuel reserves sufficient for over 20 years of operation[14], maintaining long-term instrument operability on JWST is a critical priority. Thanks to its unmatched multiplexing capability and the evolving strategies for preserving shutter functionality, the NIRSpec MSA is positioned to deliver unprecedented efficiency in the observation of distant galaxies for years to come.

**Acknowledgements**

NIRSpec was designed and built for ESA by Airbus Defence and Space GmbH in Ottobrunn, Germany, with the focal plane array and micro-shutter assembly provided by NASA Goddard Space Flight Center. We thank the dedicated engineers and scientists of ESA, Airbus, NASA, and STScI for their ongoing and enthusiastic support of NIRSpec operations.